\documentclass[11pt,a4paper]{article}

\usepackage{jheppub}
\usepackage{amsmath,amssymb,euscript,array,mathrsfs}
\usepackage{epsfig}
\usepackage{slashed}

\def\a{\alpha}
\def\b{\beta}

\def\d{\delta}

\def\l{\lambda}
\def\m{\mu}

\def\L{\Lambda}

\newcommand{\NN}{{\cal N}}
\newcommand{\MSB}{{\overline{MS}}}

\title{
\vskip2cm
The Quark Condensate in Multi-Flavour QCD -- Planar Equivalence 
Confronting Lattice Simulations}

\author[a]{Adi Armoni,} 
\author[b]{Mikhail Shifman,}
\author[a]{Graham Shore}
\author[c,d]{and Gabriele Veneziano}

\affiliation[a]{Department of Physics, Swansea University,\\
Singleton Park, Swansea, SA2 8PP, UK}

\affiliation[b]{William I. Fine Theoretical Physics Institute, University of Minnesota,\\
Minneapolis, MN 55455, USA}

\affiliation[c]{PH Department, CERN, \\
CH-1211 Geneva 23, Switzerland}

\affiliation[d]{ Coll\`ege de France, \\
11 place M. Berthelot, 75005 Paris, France}

\abstract{Planar equivalence between the large $N$ limits of 
${\cal N}=1$ Super Yang-Mills (SYM) theory and a variant of QCD with
fermions in the antisymmetric representation is a powerful tool to
obtain analytic non-perturbative results in QCD itself. 
In particular, it allows the quark condensate for $N=3$ QCD with
quarks in the fundamental representation to be inferred from exact
calculations of the gluino condensate in ${\cal N}=1$ SYM. 
In this paper, we review and refine our earlier predictions for the
quark condensate in QCD with a general number $n_f$ of flavours and
confront these with lattice results.

\vskip4cm
}

\dedicated{{\rm CERN-PH-TH-2014-255 \\
FTPI-MINN-14/43,~ UMN-TH-3414/14}}

\notoc

\begin{document}

\maketitle

\newpage

\section{The QCD condensate and planar equivalence}

One of the challenges in theoretical studies of QCD is to find
analytic, non-perturbative methods for calculations of strong-coupling
quantities such as the quark or gluon condensates. Such methods are an
important complement to direct evaluations using lattice gauge theory
and give extra physical insight into the underlying dynamical
mechanisms. 
One proposal is to use the special properties of supersymmetric
theories to perform exact non-perturbative calculations in 
${\cal N}=1$ gauge theories, then relate these in suitable limits to
infer results in QCD itself. This approach has been pioneered in
refs.\cite{Armoni:2003gp,Armoni:2004ub}.

The key idea is to exploit a remarkable property of $SU(N)$ QCD with a
single flavour of Dirac fermions in the antisymmetric
representation,\footnote{This is the theory referred to in
\cite{Armoni:2003gp,Armoni:2004ub,Armoni:2003yv}  
as ``QCD-OR'' or ``Orientifold QCD''. This name
highlights its origin in string theory \cite{Sagnotti:1995ga}, though this will play no
role in the analysis here.}
QCD$_{\rm AS}$, {\it viz.}~that as $N$ is varied, it
interpolates between three theories of special importance -- pure
Yang-Mills theory, QCD with one flavour of fundamental fermions 
and $\NN=1$ super-Yang Mills theory.

Specifically, for $N=2$, the antisymmetric representation becomes
trivial and QCD$_{\rm AS}$ becomes simply $SU(2)$ Yang-Mills.
For $N=3$, the antisymmetric representation (which has dimension
$\frac{1}{2} N(N-1)$) coincides with the fundamental representation
(dimension $N$) and so QCD$_{\rm AS}$($N=3$) is identical to
one-flavour $SU(3)$ QCD.\footnote{Here we use $n_f$ for the 
 {\it total} number of quark flavours. 
Note this is in contrast to \cite{Armoni:2005wt}
where it was used as the {\it additional} number of fundamental flavours in a
``QCD-OR$^\prime$'' theory comprising one antisymmetric flavour plus
fundamentals, which we call QCD$_{\rm AS-F}$ here.}
In the large $N$ limit, QCD$_{\rm AS}$($N\rightarrow\infty$)
becomes equivalent to a theory with $SU(N)$ gauge group and a single 
real fermion in the adjoint representation (dimension $N^2 - 1$). 
Crucially, this theory is supersymmetric, {\it viz.}~$\NN=1$ super Yang
Mills (SYM), and this is the key to being able to perform the exact
non-perturbative calculations we exploit.

The relation of QCD$_{\rm AS}$ at large $N$ with $\NN=1$ SYM has
been extensively described in a series of earlier papers on ``planar
equivalence''
\cite{Armoni:2003gp,Armoni:2004ub,Armoni:2003yv,Armoni:2005wt}. 
It has been shown that in the 't Hooft large-$N$ limit the two
theories become equivalent in the common bosonic $C$-parity even
sector. A necessary and sufficient condition for planar equivalence to
hold is that charge conjugation symmetry is not broken spontaneously 
\cite{Unsal:2006pj}. This was verified by a dedicated 
lattice simulation \cite{DeGrand:2006qb} (see also \cite{Lucini:2007as})
where it was shown that charge conjugation symmetry is broken if 
one dimension is compactified on a small-enough circle, but is 
restored at large (in particular infinite) compactification radius.

In this paper, we focus on a single issue --
the prediction of the value of the quark condensate in QCD,
its $N$ and $n_f$ dependence, and its confrontation with lattice data.
The gluino condensate \cite{Witten:1982df, Veneziano:1982ah} 
has been evaluated exactly in $\NN=1$ SYM 
\cite{Shifman:1987ia,Davies:1999uw}
and the idea here is to use QCD$_{\rm AS}$ with 
varying $N$ to infer the value of the quark condensate for one-flavour
$N=3$ QCD 
by interpolating between $\NN=1$ SYM at large $N$ and 
pure Yang-Mills at $N=2$, where of course the condensate disappears. 
For many flavours, we consider a generalisation to a theory, 
QCD$_{\rm AS-F}$, with one AS representation fermion and $(n_f-1)$
fundamentals.

The calculation of the gluino condensate 
$\langle \l\l\rangle \equiv \langle\l^{a\a}\l^a{}_\a\rangle$ 
in $\NN =1$ SYM relies on the holomorphy of $F$-terms in
supersymmetric theories to analytically continue a weak-coupling,
semi-classical evaluation of the condensate in a deformed version of
the theory to the strong-coupling regime of $\NN = 1$ SYM itself.
Specifically, in ref.\cite{Shifman:1987ia}, additional matter fields
with mass $m$ are added allowing the condensate to be calculated
from the one-instanton contribution in a weak-coupling regime 
at small non-zero $m$ before taking $m\rightarrow\infty$ to decouple
the extra fields and recover the original theory.
In ref.\cite{Davies:1999uw}, $\NN=1$ SYM itself is considered on a
compactified space $R^3 \times S$ (with $\b$ the radius of the
compactified dimension) and the condensate is evaluated initially in
the limit of small $\b$ where the theory is weakly-coupled and the
condensate is dominated by contributions from monopole configurations,
both conventional BPS type and additional Kaluza-Klein monopoles.
Both approaches agree and, quoting the result for the condensate for
an $SU(N)$ gauge group in terms of the scale $\L_\MSB$ appropriate to
SYM (see below), we have
\begin{equation}
\langle \l\l \rangle_\MSB = - ~\frac{N^2}{2\pi^2} ~\frac{3}{2\l(\m)}~
\L_\MSB^3\big|_{\rm SYM} \ .
\label{a1}
\end{equation}

\begin{table}
\centering
\begin{tabular}{c c c c c}
\hline
{} & QCD$_{\rm AS-F}(N,n_f)$ & Yang-Mills & QCD$_{\rm F}(N,n_f)$  & $\NN
= 1$ SYM \\ [0.5ex]
\hline\\[0.5ex]
%{}&{}&{}&{}&{}\\
$\b_0$ & $3N + 2 - \tfrac{2}{3}n_f$ & $\tfrac{11}{3}N$ &
$\tfrac{11}{3}N - \tfrac{2}{3}n_f$ & $3N$ \\ [0.5ex]
$\b_1$ & $3N^2 +\tfrac{17}{2}N - \frac{9}{2N} -\tfrac{1}{6}n_f(13N -
\frac{3}{N})$ 
& $\tfrac{17}{3}N^2$ & $\tfrac{17}{3}N^2 -
\tfrac{1}{6}n_f\left(13N - \frac{3}{N}\right)$ & $3N^2$ \\ [0.5ex]
$\gamma$ & $\frac{3}{N}(N-2)(N+1)~~[\rm AS]$ & -- & $\frac{3}{2N}(N^2
-1)$ & $3N$ \\ [0.5ex]
{}&$\frac{3}{2N}(N^2 -1)~~~~~[\rm F]$ &{} &{} &{} \\ [1ex]
\hline
\end{tabular}
\caption{Renormalization group coefficients for general $N$ and $n_f$  
for the theories considered here.
Gamma is the anomalous dimension for 
the condensate, {\it i.e.}~the running mass anomalous
dimension. In general, for a (complex) representation R
with $n_f$ flavours,
$\b_0 = \tfrac{11}{3}C_2(A) - \tfrac{4}{3}n_f T(R)$,
~$\b_1 = \tfrac{17}{3}C_2(A)^2 - \tfrac{1}{3}n_f T(R) \left(10 C_2(A) +
    6 C_2(R)\right)$~and $\gamma = 3 C_2(R)$, 
where $C_2(A)$ is the quadratic Casimir for the gauge group.}
\vskip0.7cm
\label{table:RG}
\end{table}

\begin{table}
\centering
\begin{tabular}{c c c c}
\hline
Representation ($R$)  & $T(R)$ & $C_2(R)$ & ${\rm dim}(R)$ \\ [0.5ex]
\hline \\ [0.5ex]
Antisymmetric ($AS$) & $\frac{1}{2}(N-2)$ & $\frac{1}{N}(N-2)(N+1)$ &
$\frac{1}{2}N(N-1)$  \\ [0.5ex]
Fundamental ($F$) & $\frac{1}{2}$ & $\frac{1}{2N}(N^2 - 1)$ & $N$ \\ [0.5ex]
Adjoint ($A$) & $N$ & $N$ & $N^2 - 1$  \\ [1ex]
\hline
\end{tabular}
\caption{The Dynkin index $T(R)$  and quadratic Casimir $C_2(R)$ 
for various representations of $SU(N)$. For a
representation $R$ of $SU(N)$ with generators $t^a$ they are defined as 
${\rm tr}~ t^a t^b = T(R) \d^{ab}$ 
and    
$\left(t^a t^a\right)_{ij} = C_2(R) \d_{ij}$
and satisfy
$T(R) {\rm dim}(A) = C_2(R) {\rm dim}(R)$. }
\label{table:group}
\end{table}

Before proceeding, we need to carefully specify our conventions and
the definitions of the key quantities used below.\footnote{Our conventions 
follow those of the Particle Data Group, QCD review, 2008
\cite{Amsler:2008zzb}. 
Since we work here with the 't Hooft
coupling $\l = g^2 N/8\pi^2$ rather than $\a_s = g^2/4\pi$, it is
more convenient to use the RG coefficients $\b_0, \b_1, \ldots$
rather than the more recent PDG 2014
\cite{Agashe:2014kda} definitions $b_0, b_1,
\ldots$ to absorb convention-dependent factors of $4\pi$.
The gluino field in (\ref{a1}) is normalised so that its kinetic term
in the SYM Lagrangian is ${\cal L} = i\bar{\lambda}\slashed{D} \lambda$.}
First, for ease of reference, in Tables \ref{table:RG} and 
\ref{table:group} we collect the $N$ and $n_f$
dependence of the main group theoretical parameters and the
renormalization group coefficients for the theories considered 
here.

Our results are presented first in terms of renormalisation group invariant
quantities, written in terms of the 't Hooft coupling.
We define the RG invariant scale parameter
\begin{equation}
\L_c ~=~ \m ~(c ~\l(\m))^{-\b_1/\b_0^2}~ e^{-N/(\b_0 \l)} \ ,
\label{a2}
\end{equation}
and the RG invariant condensate for a Dirac fermion $\psi$ as 
\begin{equation}
\langle \bar{\psi} \psi \rangle_{\tilde{c}} ~=~ \left(\tilde{c}~ \l(\m)
\right)^{\gamma/\b_0}~
\langle \bar{\psi} \psi\rangle_\MSB \ ,
\label{a3}
\end{equation}
where $\langle \bar{\psi} \psi\rangle_\MSB$ denotes the renormalised
condensate in the $\MSB$ scheme at scale $\m$.
The normalisation parameters $c$ and $\tilde{c}$ are essentially
arbitrary, but should admit an expansion in $1/N$ around a finite
$O(1)$ large-$N$ limit. This ensures the condensate matching 
condition (\ref{a5}) below is consistent with planar equivalence
at large $N$ \cite{Armoni:2003yv}.
The conventional $\MSB$ definition of the scale parameter $\L_\MSB$ is
simply $\L_c$ with $c = \b_0/2N$. Later, we will choose 
$\tilde{c} = \b_0/N$ to facilitate easy comparison with lattice results.

Expressed entirely in terms of these RG invariant quantities, the SYM gluino
condensate is therefore 
\begin{equation}
\langle \l\l\rangle_{\tilde{c} }/\L_c^3\big|_{\rm SYM} ~=~ 
- \frac{N^2}{2\pi^2} ~ c ~ \tilde{c} \ ,
\label{a4}
\end{equation}
noting that for $\NN=1$ SYM both $\gamma/\beta_0$ and
$3\beta_1/\b_0^2$ are simply 1.

\noindent\textbf{One flavour,} \textbf{QCD}$_{\textbf{AS}}$\,{\textbf :}

To determine the condensate in one-flavour QCD, we start from
the QCD$_{\rm AS}$ theory, where planar equivalence has
been firmly established.
Our basic ansatz for the QCD$_{\rm AS}$ condensate is
\begin{equation}
\langle \bar{\Psi}\Psi\rangle_{\tilde{c} }/\L_c^3\big|_{\rm AS} ~=~ 
- \frac{N^2}{2\pi^2}~\left(1 - \frac{2}{N}\right) ~ c^{3\b_1/\b_0^2} ~
\tilde{c}^{\gamma/\b_0} ~K_{\rm AS}(1/N;n_f=1)\ ,
\label{a5}
\end{equation}
where $\Psi$ denotes a fermion in the AS representation of $SU(N)$
and the appropriate RG coefficients can be read off from Table \ref{table:RG}.
The content of (\ref{a5}) is that the most significant $1/N$ correction
to the leading large $N$ behaviour of $\langle \bar{\Psi}\Psi\rangle_{\tilde{c}}$ 
as determined by planar equivalence with the exact SYM
result (\ref{a4}) is given by the relative $(1-2/N)$ factor. 
Assuming a smooth dependence of $\langle
\bar{\Psi}\Psi\rangle_{\tilde{c}}$ on $N$ in the QCD$_{\rm AS}$ theory,
this is the simplest interpolating factor between the large $N$ SYM
result and the vanishing of the condensate for $N=2$, where 
the antisymmetric representation is trivial and 
QCD$_{\rm AS}$ degenerates to pure $SU(2)$ Yang-Mills.
Notice that this factor is simply the ratio of the Dynkin indices for
the AS and adjoint representations, a feature we may conjecture 
to be more generally valid.
The remaining sub-leading corrections are encoded in the factor 
$K_{\rm AS} = 1 + O(1/N)$, which we initially assume to be
relatively small. 

 Given the arbitrariness in the normalisation of the RG invariant
 condensates and scale parameters, it is natural to separate the
 dependence on the $c$, $\tilde{c}$ factors explicitly on the rhs of
 (\ref{a5}). Notice\footnote{Explicitly, 
for a single AS representation,
$\gamma/\b_0 =  1 -\frac{13}{9} \frac{1}{N}$ and
$3\b_1/\b_0^2 = 1 + \frac{11}{9} \frac{1}{N}$, so
$\tilde{c}^{-1 + \gamma/\b_0} = 1 - \frac{13}{9N} \log{\tilde{c}} + O(1/N^2)$
and $c^{-1 + 3\b_1/\b_0^2} = 1 + \frac{11}{9N} \log{c} + O(1/N^2)$.} 
that in the {\it ratio of ratios} between (\ref{a5}) and (\ref{a4}) for
the AS and adjoint represention condensates, both these factors
are $1 + O(1/N)$ since the RG factors in
the exponents are both $O(1/N)$ (see Table \ref{table:RG}),
so could in principle be absorbed into the $K_{\rm AS}$ factor.
However, this would not be appropriate since they are clearly 
convention dependent whereas $K_{\rm AS}$ should reflect the
basic $O(1/N)$ physics of the theory. 

Our prediction for the condensate can be expressed in
several ways, which will be useful for comparing with lattice data.
In particular, we may write
\begin{equation}
\langle \bar{\Psi}\Psi\rangle_{\tilde{c} }/\L_{\MSB}^3\big|_{\rm AS} ~=~ 
- \frac{N^2}{2\pi^2}~\left(1 - \frac{2}{N}\right) ~ \left(\frac{\b_0}{2N}\right)^{3\b_1/\b_0^2} ~
\tilde{c}^{\gamma/\b_0} ~K_{\rm AS}(1/N;n_f=1)\ ,
\label{a7}
\end{equation}
and
\begin{equation}
\langle \bar{\Psi}\Psi\rangle_{\MSB}/\L_{\MSB}^3\big|_{\rm AS} ~=~ 
- \frac{N^2}{2\pi^2}~\left(1 - \frac{2}{N}\right) ~ \left(\frac{\b_0}{2N}\right)^{3\b_1/\b_0^2} ~
\l(\m)^{-\gamma/\b_0} ~K_{\rm AS}(1/N;n_f=1)\ ,
\label{a8}
\end{equation}
where in the latter form, $\langle \bar{\Psi}\Psi\rangle_{\MSB}$ is 
$\mu$-dependent and we need to find the 't Hooft coupling
$\l(\m)$ by inverting the relation (\ref{a2}) for $\L_{\MSB}$.
Finally, as used in ref.\cite{Armoni:2005wt}, we could express the
condensate entirely in terms of the 't Hooft coupling at scale $\m$,
{\it viz.} 
\begin{equation}
\langle \bar{\Psi}\Psi\rangle_{\MSB} ~=~ - \m^3 ~
\frac{N^2}{2\pi^2}~\left(1 - \frac{2}{N}\right) ~
\l(\m)^{-3\b_1/\b_0^2 - \gamma/\b_0} ~ e^{-3N/(\b_0\l(\m))} ~K_{\rm AS}(1/N;n_f=1)\ .
\label{a9}
\end{equation}

\noindent$\boldsymbol{n_f}$ \textbf{flavours,}
\textbf{QCD}$_{\textbf{AS-F}}$\,{\textbf :}

So far, we have discussed the condensate in theories with only a
single flavour, where planar equivalence with ${\cal N}=1$ SYM 
has been demonstrated for QCD$_{\rm AS}$.
In ref.\cite{Armoni:2005wt}, we explored to what extent planar
equivalence could be shown directly in a multi-flavour theory.
Since we need the additional flavours to decouple in the large-$N$ 
limit, and since we ultimately wish to discuss $N=3$ QCD with
quarks in the fundamental representation, 
we considered the hybrid theory QCD$_{\rm AS-F}$,
{\it viz.}~QCD with one AS and $(n_f-1)$ fundamental
fermions (see footnote 2). 

The demonstration of planar equivalence and matching of condensates
with ${\cal N}=1$ SYM in this case involved comparison of Wilson loops and
the construction from anomalous chiral Ward identities of a
`decoupling' current, which defines a sector in which the Goldstone
bosons of spontaneously broken chiral symmetry do not affect the
relevant correlation functions. These theoretical considerations are
described at length in \cite{Armoni:2005wt}. Here we just quote
our conclusions for the RG-invariant condensates:
\begin{equation}
\langle \bar{\Psi}\Psi\rangle_{\tilde{c} }/\bigl(\L_c^{(n_f)}\bigr)^3\big|_{\rm AS-F} ~=~ 
- \frac{N^2}{2\pi^2}~\left(1 - \frac{2}{N}\right) ~ c^{3\b_1/\b_0^2} ~
\tilde{c}^{\gamma/\b_0} ~K_{\rm AS}(1/N; n_f)\ ,
\label{a10}
\end{equation}
for the AS fermion $\Psi$, while for the fundamental fermions $q$,
\begin{equation}
\langle\bar{q} q\rangle_{\tilde{c}}/\bigl(\L_c^{(n_f)}\bigr)^3\big|_{\rm AS-F} ~=~
- \frac{N}{2\pi^2}~~ c^{3\b_1/\b_0^2} ~
\tilde{c}^{\gamma/\b_0} ~K_{\rm F}(1/N; n_f)\ ,
\label{a11}
\end{equation}
Once again, the $K$ factor for the AS representation is 
$K_{\rm AS} = 1 + O(1/N)$, with the $n_f$ dependence contained 
in the $O(1/N)$ terms. For the fundamental representation
fermions, however, we do not necessarily need to impose this. 
All that is actually required is self-consistency for $N=3$ when the
two representations coincide, {\it i.e.} $K_{\rm F}(1/3; n_f) = 
K_{\rm AS}(1/3; n_f)$.

These $K$ factors
encode the sub-dominant $1/N$ corrections, which we conjecture to be
relatively small. Our initial condensate predictions for QCD are therefore
based on taking the relevant $K \simeq 1$ and confronting
these with lattice data. Further dynamical insight and assumptions may
subsequently be used to refine the prediction. For example, in
ref.\cite{Armoni:2005wt} we used the argument that QCD with $n_f$
flavours, the $K$ factors should go {\it smoothly} to zero as the 
conformal window is approached to estimate their flavour dependence,
finding a rather mild dependence. Ideally, lattice simulations
with sufficient precision to pin down the variation of the $K$ factors
for different numbers of flavours could ultimately give information on the
behaviour of the condensate near the conformal window and the nature
of the transition.

\section{Numerical predictions and lattice data}

We now specialise to QCD with $N=3$ and $n_f$ fundamental flavours and
present numerical predictions for the $\langle \bar{q} q\rangle$
condensate based on the formulae above. These predictions will then be
critically compared with available results from lattice gauge theory.

First we need to emphasise that the result of any calculation, analytic or
lattice, is a dimensionless ratio, since the overall QCD scale is the 
free parameter of the theory. The cleanest way to present our results
is therefore in terms of the ratio $\langle
\bar{q}q\rangle_{\tilde{c}}^{1/3} / \Lambda_{\MSB}^{(n_f)}$  of the RG-invariant
condensate to the QCD scale parameter in the $\MSB$ scheme for the 
relevant number of flavours.
For ease of comparison with the lattice, we adopt here the convention
of ref.\cite{Engel:2014cka} for the RG-invariant condensate, {\it viz.}~take
$\tilde{c} = \b_0/N$. We denote this condensate by
$\Sigma_{RGI} \equiv - \langle \bar{q}q \rangle _{\tilde{c}=\b_0/N}$.
The results of the previous section imply:
%\begin{equation}
%\langle \bar{q}q\rangle_{\tilde{c}=\b_0/3} /
%\left(\Lambda_{\MSB}^{(n_f)}\right)^3  ~=~ 
%- \frac{3}{2\pi^2}~\left(\frac{\b_0}{6}\right)^{3\b_1/\b_0^2} ~
%\left(\frac{\b_0}{3}\right)^{\gamma/\b_0} ~K_{\rm F}(1/3;n_f) \ ,
%\label{b0}
%\end{equation}
%that is,
\begin{equation}
\Sigma^{1/3}_{RGI} / \Lambda_\MSB^{(n_f)}~=~
\left(\frac{3}{2\pi^2}\right)^{1/3} \,\left(\frac{\b_0}{6}\right)^{\b_1/\b_0^2} \,
\left(\frac{\b_0}{3}\right)^{\gamma/3\b_0} \,K_{\rm F}^{1/3}(1/3;n_f)\ .
\label{bb0}
\end{equation}

\begin{figure}[ht] 
\centerline{\includegraphics[width=4in]{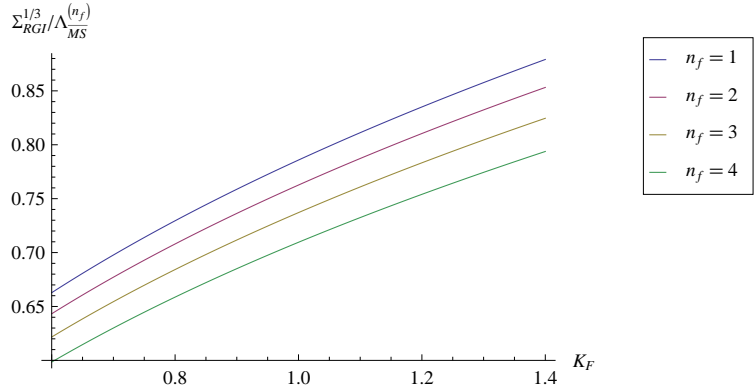}}
\caption{\footnotesize Plot showing the dependence of the RG-invariant
condensate $\Sigma_{RGI}^{1/3}/\Lambda_{\MSB}^{(n_f)}$ on the $K_F(1/3;n_f)$
parameter for different numbers $n_f$ of flavours.}
\label{Figure1}
\end{figure}

Our fundamental prediction (\ref{bb0}) is shown in Figure \ref{Figure1}, where we
plot the ratio of the RG-invariant condensate to $\Lambda_{\MSB}$ for different
numbers of flavours as a function of the $K_{\rm F}$ parameter. As we see
below, lattice data supports the view that $K_{\rm F}$ is close to 1,
so for orientation we list here our predictions taking $K_{\rm F} = 1$:
\begin{equation}
\Sigma^{1/3}_{RGI} / \Lambda_\MSB^{(n_f)}~=~  0.786~(n_f=1),~~~
0.763~(n_f=2),~~~0.737~(n_f=3),~~~0.710~(n_f=4) \ .
\label{b1}
\end{equation}

\begin{figure}[ht] 
\centerline{\includegraphics[width=4in]{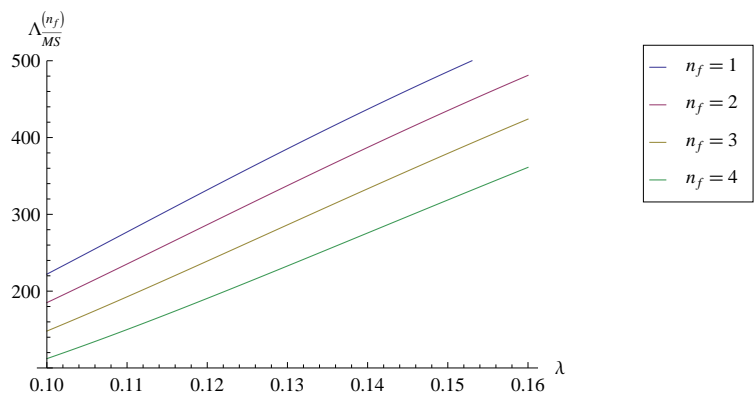}}
\caption{\footnotesize Plot showing the relation between the scale
  parameter $\Lambda_\MSB^{(n_f)}$ and the 't Hooft coupling
  $\l(\mu)$ evaluated at $\mu = 2\, {\rm GeV}$ for different numbers
  $n_f$ of flavours.}
\label{Figure2}
\end{figure}

It is also useful to express our results in terms of the $\MSB$
condensate $\Sigma_{\MSB} = \langle\bar{q} q\rangle|_{\mu=2\,{\rm GeV}}$.
This requires the relation between $\Lambda_{\MSB}$ and
the 't Hooft coupling $\l(\mu)$.
In fact, for accuracy in the numerical predictions, we do this using the
three-loop RG formula, rather than two-loop expression given above, 
{\it viz.}
\begin{equation}
\L_\MSB^{(n_f)} = \m
~\left(\frac{\b_0}{2N}\l\right)^{-\b_1/\b_0^2}~e^{-N/\b_0\l}~
\left[1 - \frac{\l}{8N}\frac{1}{\b_0^2}\left(\b_2 -
    \frac{8\b_1^2}{\b_0}\right)~ + \ldots\right] \ .
\label{bb1}
\end{equation}
This is shown, for $N=3$, in Figure \ref{Figure2}.

\begin{figure}[ht] 
\centerline{\includegraphics[width=4in]{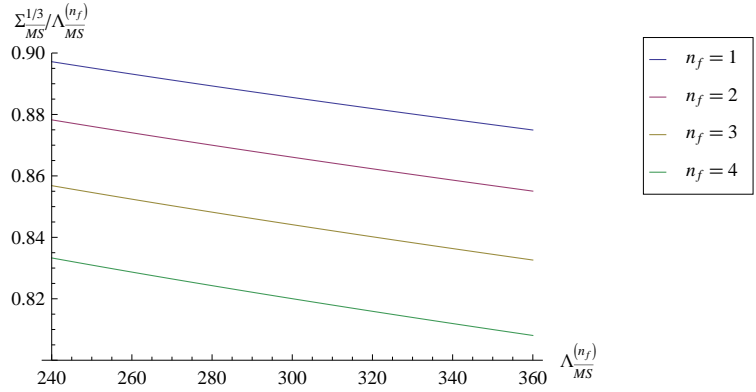}}
\caption{\footnotesize Plot showing the condensate ratio
  $\Sigma_{\MSB}^{1/3}/\Lambda_{\MSB}^{(n_f)}$ in the $\MSB$ scheme in terms of the scale
  parameter $\Lambda_{\MSB}^{(n_f)}$ for different numbers
  $n_f$ of flavours.}
\label{Figure3}
\end{figure}

\begin{figure}[ht] 
\centerline{\includegraphics[width=3in]{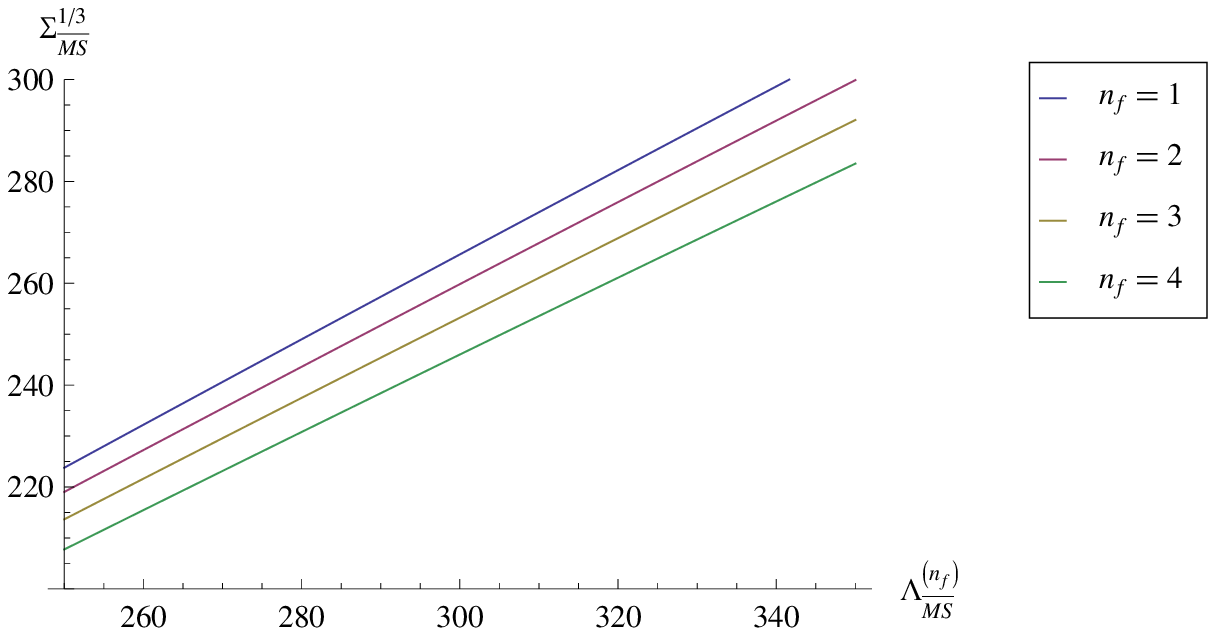} \hskip0.7cm
\includegraphics[width=3in]{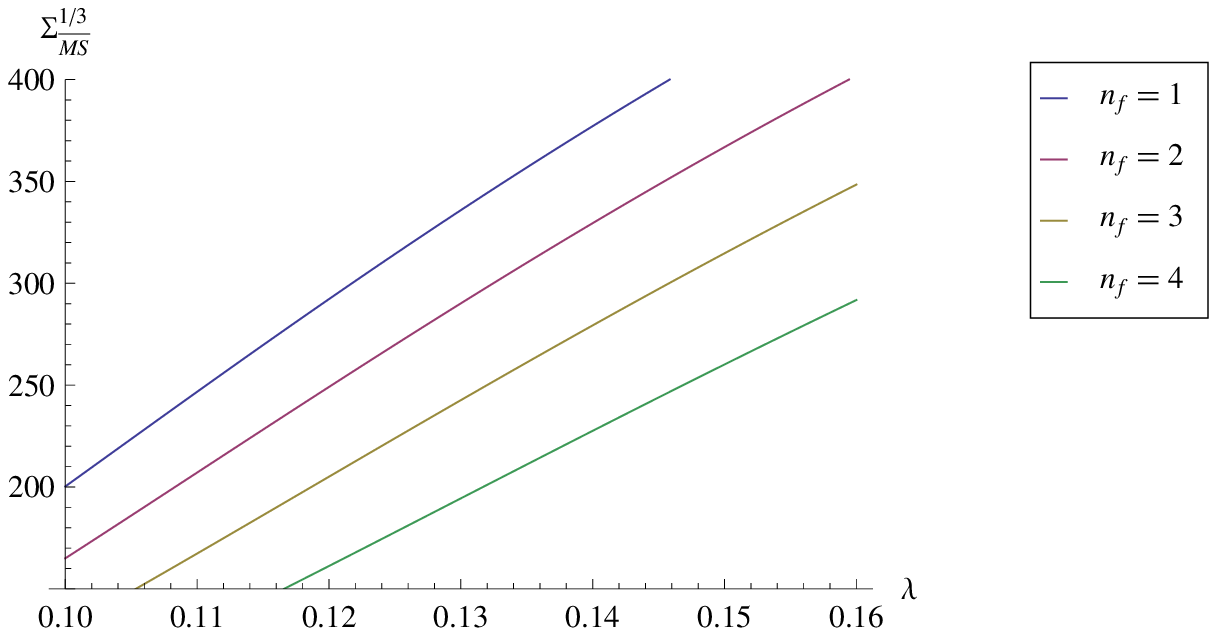}}
\caption{\footnotesize Plots showing the condensate ratio
  $\Sigma_{\MSB}^{1/3}$ in the $\MSB$ scheme at $\mu = 2\,{\rm GeV}$
  in terms of the scale parameter $\Lambda_{\MSB}^{(n_f)}$ (left) and 't Hooft coupling
  $\l(\m)$ (right) for different numbers $n_f$ of flavours.}
\label{Figure4}
\end{figure}

From the previous section, we have the following formula for the
$\MSB$ condensate in $N=3$ QCD:
\begin{equation}
\Sigma_{\MSB}^{1/3}/\Lambda_{\MSB}^{(n_f)} ~=~
\left(\frac{3}{2\pi^2}\right)^{1/3}\,
\left(\frac{\b_0}{6}\right)^{\b_1/\b_0^2} \,
\l(\mu)^{-\gamma/3\b_0}\, K_{\rm F}^{1/3}(1/3;n_f) \ .
\label{bb2}
\end{equation}
This is plotted, taking $K_{\rm F} = 1$ and evaluating at the standard
scale $\mu=2\,{\rm GeV}$, 
in Figure \ref{Figure3} for the
ratio $\Sigma_{\MSB}^{1/3}/\Lambda_{\MSB}^{(n_f)}$ and 
in Figure \ref{Figure4} for the
condensate $\Sigma_{\MSB}^{1/3}$ itself expressed in MeV units
inherited from the $n_f$-dependent scale parameter.

To confront these predictions with lattice results, we need to be
careful about interpreting scale-dependent data in variants of 
`real-world' QCD in which the number of flavours is varied. 
These are in principle distinct theories with their own independent
free scale parameter $\Lambda_{\MSB}$. Only for real-world QCD (which
we consider as $n_f=3$ light flavours with quark masses taken into
account) can predictions be unambiguously linked to experimental data, 
allowing results to be expressed in genuine MeV units.\footnote{In
  practice, a compromise is usually made whereby the MeV scale for QCD
  with $n_f \neq 3$ is set by fixing some quantity which is considered to
  be only relatively weakly dependent on $n_f$ to its experimental, real-world
  QCD, value. This is of course potentially dangerous if we are to use
  lattice results to determine the $n_f$-dependence of the condensate
  and constrain the $K_{\rm F}$ parameter.}
This means that the only strictly meaningful comparisons to be made
are between predictions of dimensionless ratios. For our purposes,
this requires comparing our predictions to a lattice calculation 
that {\it self-consistently} determines the ratio of the condensate to
$\Lambda_{\MSB}$.

While, as we discuss in the appendix, there are several evaluations in the
literature of the condensate for various $n_f$, these are usually
expressed in some definition of MeV units and are not linked to a 
self-consistent determination of $\Lambda_{\MSB}$. This makes a
precision confrontation of lattice data with our planar equivalence
predictions difficult. 

An exception is the recent work of 
Engel {\it et al.}~\cite{Engel:2014cka} and the ALPHA lattice
collaboration \cite{Fritzsch:2012wq} in which they quote
self-consistent evaluations of both the RG-invariant condensate
$\Sigma_{RGI}^{1/3}$ {\it and} the scale parameter
$\Lambda_{\MSB}^{(2)}$ for $n_f=2$. The condensate is determined by 
studying the rate of condensation of the low eigenvalues of the Dirac
operator near the limit of vanishing quark mass.
For the ratio, they quote\footnote{In ref.\cite{Engel:2014cka}, the
  results are given in terms of an auxiliary scale $F$ as
\begin{equation*}
\Sigma_{RGI}^{1/3}/F = 2.77~(2)(4), ~~~~~~~~
\Lambda_{\MSB}^{(2)}/F = 3.6~(2)
\end{equation*}
Setting MeV units by supplementing the theory with a quenched strange
quark and fixing the scale through a fit to the physical decay
constant $F_K$, they quote 
\begin{equation*}
\Sigma_{\MSB}^{1/3}\big|_{2 {\rm GeV}} ~=~ 263~(3)(4) {\rm MeV},
~~~~~~~~~\Lambda_{\MSB}^{(2)} = 311~(19) {\rm MeV}
\end{equation*}
}
\begin{equation}
\Sigma_{RGI}^{1/3} / \Lambda_{\MSB}^{(2)} ~=~ 0.77~(4)
\label{b2}
\end{equation}
Comparing with eq.(\ref{b1}) for $n_f = 2$, this is in quite
remarkable agreement with our $K_{\rm F} = 1$ prediction 
of 0.763.

\begin{figure}[ht] 
\centerline{\includegraphics[width=4in]{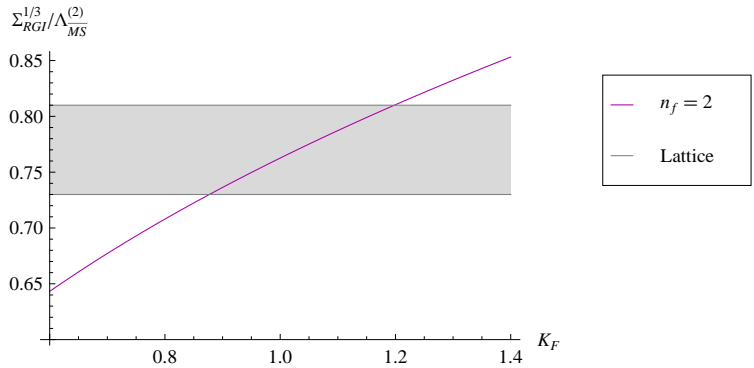}}
\caption{\footnotesize Plot showing the dependence of the RG-invariant
condensate $\Sigma_{RGI}^{1/3}/\Lambda_{\MSB}^{(n_f)}$ on the $K_F(1/3,n_f)$
parameter for $n_f=2$ superimposed with the one-sigma error band of
the lattice determination of Engel {\it et al.}}
\label{Figure5}
\end{figure}

To illustrate this further, in Figure \ref{Figure5} we restrict the plot
of $\Sigma_{RGI}^{1/3} / \Lambda_{\MSB}^{(2)}$  (see Figure
\ref{Figure1})  to $n_f =2$ and superimpose our prediction 
for the condensate as a function of $K_{\rm F}$ with the one-sigma
error band of the lattice result (\ref{b2}).
The lattice constraint on $K_{\rm F}$ is therefore
\begin{equation}
K_{\rm F}(1/3; n_f = 2) ~=~ 1.03~(16)
\label{b3}
\end{equation}
in excellent agreement with the planar equivalence prediction and our
understanding that the corrections to $K_{\rm F} \simeq 1$ are
relatively small.

Nonetheless, despite this success, it is clear that if we are to rely
on the lattice to determine $K_{\rm F}$ with the precision to gain
insight into the flavour-dependence of the quark condensate and the
transition to the conformal window, the accuracy of lattice
calculations needs to be increased, along with the extension to
self-consistent determinations of {\it both} the condensate and
$\Lambda_{\MSB}^{(n_f)}$ parameters for other values of $n_f$. 

The challenge to the lattice is therefore to extend determinations
of the quark condensate in QCD to different numbers of flavours
with the accuracy required to find a real discrimination amongst
different $n_f$. Comparison with the planar equivalence predictions
may also be stringently tested by simulations for different numbers
of colours, $N\neq 3$, or different fermion representations.
For example, in ref.\cite{Armoni:2008nq}, a lattice study of the condensate 
was carried out in the quenched approximation with fermions in the AS, 
symmetric and adjoint representations of $SU(N)$ for various values of $N$.
This broadly confirms the planar equivalence expectations and
in particular the result (\ref{a5}) that to leading order, the ratio of
condensates for different representations is given by the ratio
of their Dynkin indices.
In particular, we anticipate the following expression for the quark
condensate in a theory with fermions in the symmetric (S)
representation:
\begin{equation}
\langle \bar{\Psi}\Psi\rangle_{\tilde{c} }/\L_{\MSB}^3\big|_{\rm S} ~=~ 
- \frac{N^2}{2\pi^2}~\left(1 + \frac{2}{N}\right) ~ 
\left(\frac{\b_0}{2N}\right)^{3\b_1/\b_0^2} ~
\tilde{c}^{\gamma/\b_0} ~K_{\rm S}(1/N;n_f=1)\ .
\label{b4}
\end{equation}
Meanwhile, it would be interesting to extend the planar equivalence 
programme further by attempting analytic calculations of further
quantities beyond the gluino and quark condensates, identifying other 
scale-setting quantities more suited to comparison with the lattice 
than $\Lambda_\MSB$, and looking for further relations between 
${\cal N}=1$ SYM and QCD.

\vskip0.5cm

\centerline{*******}
\noindent We would like to thank L. Giusti and T. Hollowood for numerous discussions. 
AA and GS are grateful to the U.K.~Science and Technology Facilities Council
(STFC) for financial support under grants ST/J000043/1 and ST/L000369/1.
The work of MS is supported in part by DOE grant DE-SC0011842.

%\newpage

\appendix

\section{Lattice calculations of the condensate
  for $\boldsymbol{n_f} \mathbf = 1,2,3.$}

We present here a brief review and update of earlier exploratory
determinations of the quark condensate in $N=3$ QCD with $n_f = 1$,
2 and 3 fundamental flavours on the lattice, and their comparison with
our planar equivalence predictions. Unfortunately, this lattice data
is not sufficiently accurate to give a reliable discrimination between
different $n_f$, although as we show it agrees within its
uncertainties with our predictions.

The first comparison of the planar equivalence result with lattice
simulations was made in ref.\cite{Armoni:2003yv} with the work of DeGrand
{\it et al.}~\cite{DeGrand:2006uy}  (see also \cite{DeGrand:2006qu})
for $n_f=1$. Scale-setting in the $n_f=1$ theory was performed in
\cite{Armoni:2003yv} by equating $\Lambda_{\MSB}^{(1)}$ with the value
of $\Lambda_{\MSB}^{(3)}$ inferred from the experimental value of
$\l(\mu = 2\,{\rm GeV})$ in physical $n_f = 3$ QCD
to obtain a prediction in MeV units. However, this does does not
correspond with the scale-setting used in the lattice calculation.
Here, we improve on this comparison and update the
result of \cite{DeGrand:2006uy} using more recent lattice data
for the scales involved. 

The essential result of \cite{DeGrand:2006uy} is a value for the
$\MSB$ condensate at $\mu = 2\, {\rm GeV}$ in units of the Sommer
parameter $r_0$, {\it viz.} $r_0 \Sigma_{\MSB}^{1/3} = 0.68\,(2)$.
The scale $\Lambda_{\MSB}$ was introduced using the then current values
of the ALPHA collaboration \cite{DellaMorte:2004bc}, {\it viz.}
$r_0 \Lambda_{\MSB}^{(0)} = 0.60\,(8)$ and 
$r_0 \Lambda_{\MSB}^{(2)} = 0.62\,(6)$
with $r_0 \simeq 0.5\,{\rm fm} \simeq  (400\,{\rm MeV})^{-1}$, corresponding 
within errors to an approximately $n_f$-independent value
taken as $\Lambda_\MSB^{(2)} = 245\,(20)\,{\rm MeV}$.
We can, however, improve on this if we take the most recent ALPHA
determination of $\Lambda_\MSB^{(2)}$ from \cite{Fritzsch:2012wq}
and, still assuming $r_0 \Lambda_{\MSB}^{(n_f)}$ is not too sensitive
to $n_f = 1$ or $2$, use this to set the scale for the DeGrand {\it et
  al.}~calculation. We therefore take $r_0 = 0.503\,(10)\, {\rm fm}$ and
$r_0 \Lambda_{\MSB}^{(2)} = 0.78\,(6)$,
corresponding to $\Lambda_\MSB^{(2)} = 310\,(20)\, {\rm MeV}$
\cite{Fritzsch:2012wq}, and combining this with the value of 
$r_0 \Sigma_{\MSB}^{1/3}$ given above, we now deduce
\begin{equation}
\Sigma_{\MSB}^{1/3} / \Lambda_{\MSB}^{(1)}  ~=~ 0.87\,(7) \ ,
\label{A1}
\end{equation}
and $\Sigma_{\MSB}^{1/3} = 270\,(20) \,{\rm MeV}$.
This is to be compared with the $K_{\rm F} = 1$ planar equivalence prediction
for $n_f=1$ (see Figure \ref{Figure3})
\begin{equation}
\Sigma_{\MSB}^{1/3} / \Lambda_{\MSB}^{(1)}  ~=~ 0.884   \ ,  
\label{A2}
\end{equation}
corresponding to 
$\Sigma_{\MSB}^{1/3} = 274\,{\rm MeV}$. With this improved
scale-setting, we see that the $n_f=1$ lattice result is indeed now in
good agreement, within its significant uncertainty, with the planar
equivalence prediction.

A similar improvement  can be applied to the original $n_f=2$ condensate
prediction by DeGrand {\it et al.}~in ref.\cite{DeGrand:2006nv}.
Taking the result\footnote{In fact, ref.\cite{DeGrand:2006nv} quotes
  three values for the condensate corresponding to simulations with
  different quark masses. Two of these are in close agreement, while
  the third, for the lightest quark mass, is substantially higher and is
  disregarded in the average quoted above.}
given there as $r_0 \Sigma_{\MSB}^{1/3} = 0.69\,(2)$, we find
$\Sigma_{\MSB}^{1/3} / \Lambda_{\MSB}^{(2)}  ~=~ 0.88\,(7)$.
This is to be compared with the $K_{\rm F}=1$ planar equivalence prediction $0.864$
(see Figure \ref{Figure3}) which, with $\Lambda_{\MSB}^{(2)} = 311\,{\rm
  MeV}$ \cite{Engel:2014cka}, corresponds to 
$\Sigma_{\MSB}^{1/3}  = 269\,{\rm MeV}$.
Again we recover reasonable agreement, bringing the result of
ref.\cite{DeGrand:2006nv} into line with the precision
calculation of Engel {\it et al.}~\cite{Engel:2014cka}, 
for which this ratio is $0.85\,(5)$. 

For $n_f = 3$, our planar equivalence prediction is 
\begin{equation}
\Sigma_{\MSB}^{1/3} / \Lambda_{\MSB}^{(3)}  ~=~ 0.839   \ .
\label{A3}
\end{equation}
If we set the scale by using the PDG \cite{Agashe:2014kda} value 
for the 't Hooft coupling $\l(\mu=2\, {\rm GeV}) = 0.143$, 
corresponding to $\Lambda_{\MSB}^{(3)} = 339\,(10)\, {\rm MeV}$
our $K_{\rm F} = 1$ prediction is 
$\Sigma_{\MSB}^{1/3} =  284\,{\rm MeV}$.
This is again supported by recent lattice results, taking {\it
  e.g.} $\Sigma_{\MSB}^{1/3} =  283\,(2)\, {\rm MeV}$ ~\cite{McNeile:2012xh}
as a representative figure.

\end{document}